\documentclass[a4paper,11pt]{article}
\usepackage{jheppub,graphicx,floatflt}
\usepackage{amsmath,latexsym,amssymb,slashed}
\usepackage{enumerate}
\usepackage{mathrsfs}
\usepackage[colorlinks=true,%
            linkcolor=blue,%
            citecolor=blue,%
            urlcolor=blue]{hyperref}

\usepackage{graphicx} 
\usepackage{adjustbox} 
\usepackage{amsfonts}
\usepackage{amsmath} 
\usepackage{amssymb} 
\usepackage{mathabx} 
\usepackage{float}
 
\usepackage{subcaption}

\usepackage{tipa}
\usepackage{textcomp}
\usepackage[T1]{fontenc} 
\usepackage[utf8]{inputenc}

\usepackage{tabularx} 

\usepackage{psfrag} 

\usepackage{dsfont}
\usepackage{multirow}

\usepackage{tikz}
\usepackage[compat=1.1.0]{tikz-feynman}

\newcommand\beal{\begin{align}}

\newcommand{\eq}[1]{\begin{equation}#1\end{equation}}
\newcommand{\spl}[1]{\begin{split}#1\end{split}}

\newcommand{\beq}{\begin{equation}}
\newcommand{\eeq}{\end{equation}}
\def\bea#1\eea{\begin{align}#1\end{align}}
\def\beal#1\eeal{\begin{subequations}\begin{align}#1\end{align}\end{subequations}}

\renewcommand{\i}{\ensuremath{\textnormal{i}}}

\def\d{\text{d}}

\def\cf{{\it cf.}}

\def\slashchar#1{\setbox0=\hbox{$#1$}           
\dimen0=\wd0                                 
\setbox1=\hbox{/} \dimen1=\wd1               
\ifdim\dimen0>\dimen1                        
\rlap{\hbox to \dimen0{\hfil/\hfil}}      
#1                                        
\else                                        
\rlap{\hbox to \dimen1{\hfil$#1$\hfil}}   
/                                         
\fi}

\renewcommand{\i}{\ensuremath{\textnormal{i}}}

\usepackage{xcolor}
\usepackage{color}

\newcommand{\cnote}[1]{}

\title{Relative scale separation in orbifolds of $S^2$ and $S^5$}

\author{Dimitrios Tsimpis}

\affiliation{
Institut de Physique des Deux Infinis de Lyon \\
Universit\'{e} de Lyon, UCBL, UMR 5822, CNRS/IN2P3 \\
4 rue Enrico Fermi, 69622 Villeurbanne Cedex, France}

\emailAdd{tsimpis@ipnl.in2p3.fr}

\abstract{
In  orbifold vacua  containing an 
 $S^q/\Gamma$ factor, we compute the relative order of scale separation, $r$, 
 defined as the ratio of the  eigenvalue of the lowest-lying $\Gamma$-invariant state of the scalar Laplacian on  $S^q$, to the eigenvalue of 
 the lowest-lying  state.~For $q=2$  and $\Gamma$ finite subgroup of $SO(3)$,    or  $q=5$ and  
$\Gamma$ finite subgroup of  $SU(3)$, 
the maximal relative order of scale separation that can be achieved  is $r=21$ or $r=12$, respectively. 
For smooth $S^5$ orbifolds, the maximal relative scale separation is $r=4.2$.~Methods from invariant theory are  
very efficient in constructing $\Gamma$-invariant spherical harmonics, and can be readily generalized to other orbifolds.
}

\begin{document}
\maketitle
\flushbottom
\setcounter{footnote}{0}
\renewcommand{\thefootnote}{\arabic{footnote}}
\setcounter{section}{0}

\section{Introduction and summary}

Einstein $(p+q)$-dimensional gravity  minimally coupled to a $q$-form,     
admits Freund-Rubin solutions \cite{Freund:1980xh} of the form AdS$_p\times M_q$, with $M_q$ a $q$-dimensional Einstein manifold of positive curvature.~When embedded in supergravity theories arising from low-energy limits of string theory,~these solutions 
exhibit what is called an absence of  scale separation:\footnote{Note however that for $p$, $q$ unconstrained, these Freund Rubin solutions exhibit parametric scale separation: $L_{\text{AdS}} / L_{\text{int}}\rightarrow\infty$, for $p/q\rightarrow\infty$, {\it cf.}~\eqref{a7}.
The large-dimension behavior of these solutions was recently discussed in \cite{Bonnefoy:2020uef} in the context of the swampland conjectures.}~$L_{\text{AdS}} \sim L_{\text{int}}$ 
where  $L_{\text{AdS}}$, $L_{\text{int}}$ is the radius of curvature of AdS$_p$, $M_q$ respectively, see \S\ref{app:conv} for our conventions.

More generally, for any compact internal manifold $M_q$, the order of scale separation of an AdS$_p\times M_q$ solution can be defined as,
\eq{\label{ssdef}
r_0\equiv  L_{\text{AdS}}^2 m^2_{\text{KK}}
~,}
where $m^2_{\text{KK}}$ is the lowest eigenvalue of minus the scalar Laplacian on $M_q$, which is also identified with the square of the lowest-lying Kaluza-Klein (KK) mass.~Thus the absence of scale separation  can be stated as 
$r_0=\mathcal{O}(1)$.

Other known classes of  (possibly warped) AdS$_p\times M_q$ solutions, with more complicated form profiles and internal geometries $M_q$, can have different values of $r_0$, although the absence of scale separation persists in large classes of ``pure'' supergravity solutions \cite{Tsimpis:2012tu,Richard:2014qsa,Lust:2020npd, DeLuca:2021mcj}, {\it i.e.}~in the absence of  sources such as orientifolds.\footnote{In the case of warped products of the form AdS$_p\times M_q$, the KK square masses are given as eigenvalues of 
a certain modified Laplacian operator on $M_q$ \cite{Csaki:2000fc,Bachas:2011xa}.}~In  \cite{Gautason:2015tig},~a different definition of scale separation was used\footnote{Although different definitions of the order of scale separation are possible, this does not affect the results of the present note, which focuses on the order of relative scale separation, {\it cf.}~\eqref{1}.~In particular,  instead of \eqref{ssdef}, \cite{Gautason:2015tig} uses the quantity 
$L_{\text{AdS}}^2 m^4_{\text{KK}}$ as the order of scale separation, measured in units of the 
lower-dimensional Planck mass.~As was shown in \cite{Junghans:2020acz}, the bound derived in \cite{Gautason:2015tig} does not imply absence  of parametric scale separation, defined as the condition: $r_0\rightarrow c$ as  $L_{\text{AdS}}\rightarrow\infty$.}, that was argued to be excluded under the same assumptions as in  the no-go theorems against de Sitter \cite{Gibbons1,Gibbons:2003gb,Maldacena:2000mw}.  In  \cite{Junghans:2020acz} it was shown that, in the absence of orientifolds, for every AdS solution of type II supergravity,  there exists a certain parametric limit for which $L_{\text{AdS}}\rightarrow\infty$, $r_0\rightarrow c$, with $c$ a positive constant.\footnote{This is the same scaling as in the strong AdS distance conjecture \cite{Lust:2019zwm}, however the argument in \cite{Junghans:2020acz} does not rely on supersymmetry.} 

In the presence of orientifolds, the  vacua of \cite{DeWolfe:2005uu} appear to enjoy parametric scale separation. 
As was pointed out in \cite{Acharya:2006ne},  the  ten-dimensional supergravity description of DGKT involves smeared O6 planes, and belongs  
to the general class of $\mathcal{N}=1$ AdS$_4$ $SU(3)$-structure compactifications of massive IIA  \cite{Lust:2004ig}. A double T-duality then gives  
an AdS$_4$ solution on the Iwasawa manifold (or a certain T$^2$ fibration over K3) \cite{Caviezel:2008ik}, which was recently argued to also enjoy 
scale separation \cite{Cribiori:2021djm}.  The DGKT vacua have been criticized, not least for their use of smeared orientifolds \cite{Banks:2006hg,McOrist:2012yc}.
Recently, localized versions of DGKT have been constructed, to leading order in a certain large-flux expansion in \cite{Junghans:2020acz, Marchesano:2020qvg}. 
The question of scale separation has been revived within the context of the swampland program \cite{Lust:2019zwm,Font:2019uva, Blumenhagen:2019vgj,Baume:2020dqd,Buratti:2020kda,Lavdas:2020tyd,Cribiori:2021djm, Castellano:2021mmx}. In particular, the strong AdS distance conjecture would imply that the  DGKT vacua  
are inconsistent in a theory of quantum gravity.

Starting from a Freund-Rubin solution  of the form AdS$_p\times M_q$  without scale separation,  $L_{\text{int}} \sim L_{\text{AdS}}$, one may try to reduce the size of $M_q$ by orbifolding by a discrete group $\Gamma$. Naively, one might estimate  $L_{\text{int}} \sim V^{1/q}$, where $V$ is the volume of $M_q$, so that 
orbifolding would reduce the radius of curvature by a factor of $|\Gamma|^{1/q}$, where $|\Gamma|$ is the order of $\Gamma$.  
Moreover, suppose that $\Gamma$ belongs to 
an infinite series of finite groups parameterized by $n\in\mathbb{N}$, such that   $|\Gamma|\rightarrow\infty$ for $n\rightarrow\infty$.~The orbifolding would then seem to offer a straightforward way of obtaining scale-separated vacua. 

This naive expectation fails for  two reasons.~Firstly,  the orbifolding can reduce the 
size of $M_q$ below the curvature radius of  AdS$_p$,  but only  in a subset of the directions:~it is expected that some directions of $M_q$ remain of the order of $L_{\text{AdS}}$, even after orbifolding, excluding parametric scale separation \cite{Polchinski:2009ch}. This phenomenon can be seen very explicitly in the $S^2/\mathbb{Z}_n$ orbifold discussed below:~while orbifolding decreases the size of the longitudinal direction of $S^2$ by a factor of $n$, the size of the latitudinal direction remains unchanged,  {\it cf}.~fig.~\eqref{fig:o1} below. 
The other reason is that the action of $\Gamma$ on $M_q$ may have fixed points, in which case the orbifold $M_q/\Gamma$ has singularities, invalidating its description 
in the supergravity approximation.~In certain cases a singular  $M_q/\Gamma$ admits a well-defined description within string theory, however the string-theory description 
includes new light states localized at the singularities, so that   not all orbifold modes can be decoupled in the $|\Gamma|\rightarrow\infty$ limit.

The main focus of this note is on orbifolds of the form $S^q/\Gamma$, with $\Gamma\subset SO(q+1)$. 
It is useful to think of $S^q$ as embedded in an ambient $\mathbb{R}^{q+1}$ space, so that the action of $\Gamma$ on $S^q$ is the same as its action on the 
angular coordinates of $\mathbb{R}^{q+1}$ (parameterized in spherical coordinates).~If the action of $\Gamma$ only has the origin of $\mathbb{R}^{q+1}$ as a fixed point, 
the $S^q/\Gamma$ orbifold is smooth, otherwise there are orbifold singularities.\footnote{The singular locus of an orbifold is the set of points $p$ whose stabilizers $\Gamma_p\subseteq \Gamma$ are  non-tivial   \cite{Thurston_orbifolds}. Of course  $\Gamma\subset SO(q+1)$ stabilizes  the origin of $\mathbb{R}^{q+1}$, but the latter does not belong to $S^q$.}  Since the orbifolding does not change the local properties of the geometry  away from the singularities, the orbifolded geometry remains a local solution of (super)gravity, although the orbifolding may break all or part of the supersymmetry the original solution may possess.  

Although singular orbifold vacua   are not globally well-defined in supergravity, 
they can be  well-defined in string theory \cite{Dixon:1985jw,Dixon:1986jc}. 
In certain cases an AdS$_p\times S^q/\Gamma\times M_d$  supergravity solution can be thought of as the near-horizon limit of string-theory backgrounds with branes embedded in 
$\mathbb{R}^{1,p-2}\times\mathbb{R}^{q+1}/\Gamma\times M_d$, which can be studied with perturbative string-theory methods  \cite{Douglas:1996sw,Douglas:1997de,Lawrence:1998ja,Hanany:1998sd}. 

Prominent examples of sphere orbifolds can be obtained starting from  
the $\text{AdS}_5\times S^5$ vacua of IIB and the  $\text{AdS}_2\times S^2\times$ CY$_3$ vacua of IIA (the CY$_3$ can also be replaced by 
T$^6$, or K3$\times$T$^2$). As is well known, the  former can be viewed as the near-horizon geometry of a stack of D3 branes in flat space, sitting at the origin of the transverse $\mathbb{R}^6$ \cite{Maldacena:1997re}, 
while the latter can be viewed as the near-horizon geometry of a stack of D0 and three stacks of intersecting D4 branes wrapping four-cycles of the CY$_3$ and sitting at the origin of the transverse $\mathbb{R}^3$, see \cite{Bonnefoy:2019nzv,Lust:2020npd} for a recent discussion. In either case, representing the transverse space to the branes, $\mathbb{R}^{q+1}$, as a cone over $S^q$, we may orbifold by a subgroup $\Gamma\subset SO(q+1)$ acting on the base of the cone. This results in a near horizon geometry of the form  $\text{AdS}_5\times S^5/\Gamma$   \cite{Kachru:1998ys}  and  $\text{AdS}_2\times S^2/\Gamma\times$ CY$_3$, for 
$q=5,2$ respectively.

In this case the  string theory  spectrum consists of the untwisted and twisted sectors. The former is  the projection of the spectrum of the unorbifolded theory to 
the $\Gamma$-invariant states (which includes the $\Gamma$-invariant supergravity states as a subset). 
When the discrete group is freely-acting, the masses of all twisted-sector states scale as $L_S/\alpha'$ (as they correspond to strings stretching between different points of $S^q$ that are identified under the action of $\Gamma$). In the supergravity regime, $L_S^2\gg \alpha'$, the twisted-sector  states are thus much heavier than the KK mass, which scales 
as $1/L_S$.~However, if the action of $\Gamma$ has fixed points on $S^q$, the twisted sector includes light states localized at the singular locus, which must be taken into account 
in the low-energy description of the theory, in addition to the $\Gamma$-invariant supergravity fields.~Thus, 
as already mentioned, in the case of singular orbifolds, the orbifolding does not increase the order of scale separation.

In the case of smooth orbifolds, in a regime where the supergravity low-energy description  is valid, in 
order to determine the  scale separation after orbifolding,  we must examine the $\Gamma$-invariant states of the spectrum of the scalar Laplacian on $S^q$. 
  Let $m^2_{\text{KK}}$ and $m^2_{\text{KK}/\Gamma}$ 
be the lowest eigenvalue of (minus) the scalar Laplacian spectrum on  $S^q$ and its $\Gamma$-invariant projection respectively.  
The relative order of scale separation is determined by the ratio,  
\eq{\label{1}
r\equiv \left(\frac{m_{\text{KK}/\Gamma}}{m_{\text{KK}}}\right)^2
~, }
so that the total  scale separation after orbifolding is given by $r_0r$, {\it cf.}~\eqref{ssdef}. 

To our knowledge, 
a systematic calculation of $r$, for different discrete groups $\Gamma$, is not available in the literature.~Although, 
as already mentioned, we do not expect the orbifolding to lead to parametric scale separation, $r\rightarrow\infty$, 
the exact value of $r$ might be of interest in applications involving effective 
field theories in AdS space, see {\it e.g.}~\cite{Costantino:2020msc} for a recent discussion. 
In the present note we calculate the 
relative order of scale separation $r$  for the $q=2,5$ cases, for all $\Gamma\subset SO(3)$ and $\Gamma\subset SU(3)$ respectively.

Non-supersymmetric  $\text{AdS}_5\times S^5/\Gamma$  vacua have been shown to be perturbatively unstable  if $\Gamma$ has fixed points on $S^5$  \cite{Dymarsky:2005nc}.   
If $\Gamma$ is freely-acting, these vacua can be perturbatively stable, however they are 
unstable non-perturbatively \cite{Horowitz:2007pr} against tunneling to a Witten-type bubble of nothing \cite{Witten:1981gj}.~These results have been generalized to other 
non-supersymmetric AdS$_5$ backgrounds in \cite{Martin:2008pf, Ooguri:2017njy}. More generally, it has been conjectured that all non-supersymmetric AdS vacua supported by fluxes are unstable \cite{Ooguri:2016pdq, Freivogel:2016qwc, Danielsson:2016mtx}, and recent evidence from the study of concrete examples does not contradict the conjecture \cite{Basile:2021vxh, Marchesano:2021ycx, Junghans:2022exo}. Since it is expected that supersymmetry is broken unless $\Gamma\subset SU(3)$, in this note  we will  focus on this case.

Our results for $r$ as a function of $\Gamma$ are summarized in the two tables \ref{tab:1} and \ref{tab:2}.~As anticipated, we confirm the absence of parametric scale separation:~the maximal value of $r$ that can be achieved is $r=21$ and $r=12$, for the case of $S^2/\Gamma$ and $S^5/\Gamma$ respectively. Interestingly, although both   
$SO(3)$ and $SU(3)$ contain finite subgroups that come in infinite series, the maximal relative scale separation is achieved when $\Gamma$ belongs to one of the  exceptional subgroups.~All of the $S^5/\Gamma$ orbifolds admit Killing spinors, and are thus expected to lead to a supersymmetric theory on AdS$_5\times S^5/\Gamma$. 
In contrast, none of the $S^2/\Gamma$ admit Killing spinors, thus leading to a non-supersymmetric theory on $\text{AdS}_2\times S^2/\Gamma\times$ CY$_3$. All of the $S^2$ orbifolds are singular, while the only smooth $S^5/\Gamma$ orbifolds are obtained for $\Gamma=\mathbb{Z}_n$ or $T_n$. In this case the relative scale separation is $r=2.4$ and $r=4.2$ respectively.
\begin{center}
\captionof{table}{Orbifolds of $S^2/\Gamma$ for finite groups $\Gamma\subset SO(3)$ of order $|\Gamma|$, listed with their corresponding relative order of scale separation $r$, 
and the lowest degree  
$k$ for which there is a $\Gamma$-invariant spherical harmonic on $S^2$.~The explicit form of the invariants is given in \S\ref{sec:2}.~None of these orbifolds admits Killing spinors; all of them contain singular points.\\}\label{tab:1}
\begin{tabular}{|c || c|c|c|c|c| }
  \hline
$\Gamma$ & $\mathbb{Z}_n$ & $\mathbb{D}_n$ & $\mathcal{T}$ & $\mathcal{O}$ & $\mathcal{I}$  \\
  \hline
  $|\Gamma|$ & $n$ & $2n$ & $12$ & $24$ & $60$  \\
  \hline
  $k$ & 1 & 2 &3&4&6 \\
   \hline
  $r$ & 1 & 3 &6&10&21 \\
  \hline
\end{tabular}
\end{center}
\begin{center}
\captionof{table}{Orbifolds of $S^5/\Gamma$ for finite groups $\Gamma\subset SU(3)$ of order $|\Gamma|$, listed with their  corresponding relative order of scale separation $r$, 
and the lowest degree  $k$ for which there is a $\Gamma$-invariant spherical harmonic on $S^5$.~The explicit form of the invariants is given in \S\ref{sec:4}.~All of these orbifolds admit Killing spinors; only the $T_n$, $\mathbb{Z}_n$ orbifolds are smooth.\\}\label{tab:2}
\begin{tabular}{|c || c|c|c|c|c| }
  \hline
$\Gamma$  & $\Gamma_{U(2)}$ & $\mathbb{Z}_m\times \mathbb{Z}_n$  & $T_n$ & $\Delta(3n^2)$ & $\Delta(6n^2)$  \\
  \hline
  $|\Gamma|$ & $|\Gamma_{U(2)}|$  & $m\times n$  & $3n$ & $3n^2$ & $6n^2$  \\
   \hline
  $k$ & 2 & 2 &3&$\leq 3$&$\leq 4$ \\
  \hline
  $r$ & 2.4 & 2.4 &4.2&$\leq4.2$&$\leq6.4$ \\
  \hline
\end{tabular}
\vskip .5cm
\begin{tabular}{|c || c|c|c|c|c| c|c|c|}
  \hline
$\Gamma$ &   $\Sigma(60)$ & $\Sigma(60)\times\mathbb{Z}_3$ & $\Sigma(168)$ & $\Sigma(168)\times\mathbb{Z}_3$ &$\Sigma(36\varphi)$ & $\Sigma(72\varphi)$ & $\Sigma(216\varphi)$ & $\Sigma(360\varphi)$ \\
  \hline
  $|\Gamma|$ &   $60$ & $180$ & $168$ & $504$ &$108$ & $216$ & $648$ & $1080$ \\
\hline
  $k$ & 2 & 4 &4&6&4 &5&6&6 \\
   \hline
  $r$ & 2.4 & 6.4 &6.4&12&6.4 &9&12&12 \\
  \hline
\end{tabular}
\end{center}

\vskip .3cm

The outline of the rest of the paper is as follows. In \S\ref{sec:2} we present the $S^2/\Gamma$ orbifolds and calculate the relative order of scale separation for each 
$\Gamma\subset SO(3)$. 
We  do so first by using brute force, by determining the lowest degree 
$k$ for which there is a $\Gamma$-invariant spherical harmonic on $S^2$, {\it cf.}~\S\ref{app:a}. The relative order of scale separation can then be read off of \eqref{a8}. 
This procedure becomes cumbersome rather quickly, so a more efficient method is needed in order to tackle the orbifolds of $S^5$. 
The relevant tools are provided by invariant theory, and are presented in \S\ref{sec:molien}. 
Equipped with this technology, we return to the $S^2$ quotient by the icosahedral group in  \S\ref{sec:s2I2}, and calculate $r$ in an alternative way, 
using methods from invariant theory. In \S\ref{sec:4} we present the $S^5/\Gamma$ orbifolds, and calculate the relative order of scale separation for each $\Gamma\subset SU(3)$. 
In \S\ref{sec:smooth} we discuss the smoothness of the orbifolds, and the existence of Killing spinors. Several technical details can be found in the appendices. 

\noindent {\bf Note added:} After this work was completed, I became aware of the recent paper \cite{Collins:2022nux} which also contains the results for the smooth $S^5/\Gamma$ orbifolds presented here (cases $\Gamma=\mathbb{Z}_n,T_n$ of Table 2).


\section{Orbifods of $S^2$}\label{sec:2}

The  orbifolds of $S^2$ are obtained by orbifolding by a finite subgroup $\Gamma\subset SO(3)$, viewed as acting on the coordinates of the ambient $\mathbb{R}^3$. 
In the case of orientation-preserving (also known as ``chiral'' or ``proper'') orbifolds, 
$\Gamma$ can be obtained from the finite subgroups of $SU(2)$, {\it cf.}~\S\ref{app:su2}, via the 2:1 map $SU(2)\rightarrow SO(3)$. They consist of 
two infinite series and three exceptional cases, corresponding to the symmetry groups of the Platonic solids:

\begin{itemize}

\item The cyclic groups $\mathbb{Z}_{n}$, $n\geq2$, of order $n$, generated by $R_z( \frac{2\pi}{n})$, where 
$R_{\vec{u}}(\theta)$ denotes a rotation of angle $\theta$ around the axis $\vec{u}$.

\item The dihedral groups $D_{n}$, $n\geq2$, of order $2n$,  generated by $R_z( \frac{2\pi}{n})$ and $R_x( \pi)$. 

\item The tetrahedral group $\mathcal{T}$ of order 12 is isomorphic to $A_4$, the set of even permutations of four objects.~It is obtained  by combining $D_2$, 
which is generated by $R_z(\pi)$ and $R_x( \pi)$, with the element, 
\eq{
\left( {\begin{array}{ccc}
0 &1 &0\\
  0 & 0&1\\
   1&0&0
  \end{array} } \right)
~,}
which generates cyclic permutations of the coordinates of $\mathbb{R}^3$. 

\item The octahedral group $\mathcal{O}$ of order 24 is isomorphic to $S_4$, the set of permutations of four objects. It is obtained  by combining $\mathcal{T}$ 
with $R_z( \frac{\pi}{2})$.

\item The icosahedral group $\mathcal{I}$ of order 60  is isomorphic to $A_5$, the set of even permutations of five objects. It is generated by  $R_z( \frac{2\pi}{5})$ and,
\eq{\label{2.2}R=
\left( {\begin{array}{ccc}
\frac{1}{\sqrt{5}} &0 &\frac{2}{\sqrt{5}}\\
  0 & -1&0\\
   \frac{2}{\sqrt{5}}&0&-\frac{1}{\sqrt{5}}
  \end{array} } \right)
~.}

\end{itemize}
In the following we will calculate the first non-zero scalar Laplacian eigenvalue for each of these orbifolds. In the present section, we will do so by using brute force, {\it i.e.}~by determining the lowest degree  
$k$ for which there is a $\Gamma$-invariant spherical harmonic on $S^2$, {\it cf.}~\S\ref{app:a}. The relative order of scale separation can then be read off of \eqref{a8}.~A more sofisticated method, using invariant theory, will be presented in the following sections.

\subsection{$S^2/\mathbb{Z}_n$}

In the case of the cyclc groups $\Gamma=\mathbb{Z}_n$, the fundamental domain of the orbifold is the slice of the sphere bounded by the two meridians at $\phi=0$ 
and $\phi=\frac{2\pi}{n}$, with two conical singularities of degree $n$  at the poles. We see that 
while orbifolding decreases the size of the longitudinal direction by a factor of $n$, the size of the latitudinal direction remains unchanged, as depicted in fig.~\eqref{fig:o1}. 
This is also reflected in the fact that the lowest non-trivial scalar Laplacian eigenvalue occurs at degree $k=1$, just as in the unorbifolded case, {\it cf}.~\S\ref{app:a}. To see this, 
it suffices to note that the polynomial $p_{k=1}(\vec{x})=z$, {\it cf}.~\S\ref{app:a1}, is invariant under the orbifolding. We thus obtain $r=1$, i.e.~no relative scale separation in this case.

\subsection{$S^2/D_n$}

The $S^2/D_n$ orbifold can be thought of as modding out 
$S^2/\mathbb{Z}_n$ by a parity transformation in the $y$, $z$ coordinates, i.e.~$\theta\rightarrow\pi-\theta$, $\phi\rightarrow2\pi-\phi$ in 
spherical coordinates. This is half the fundamental domain of  $S^2/\mathbb{Z}_n$, since the southern half of the slice bounded by the 
$\phi=0, \frac{2\pi}{n}$ meridians is identified with the northern half. 
We now have three conical singularities: two of degree two at $\phi=0, \frac{2\pi}{n}$ and $\theta=\frac{\pi}{2}$,  and a third singularity of degree $n$ at the north pole. 

\begin{figure}[H]
\begin{center}
\begin{subfigure}[H]{0.4\textwidth}
\includegraphics[width=\textwidth]{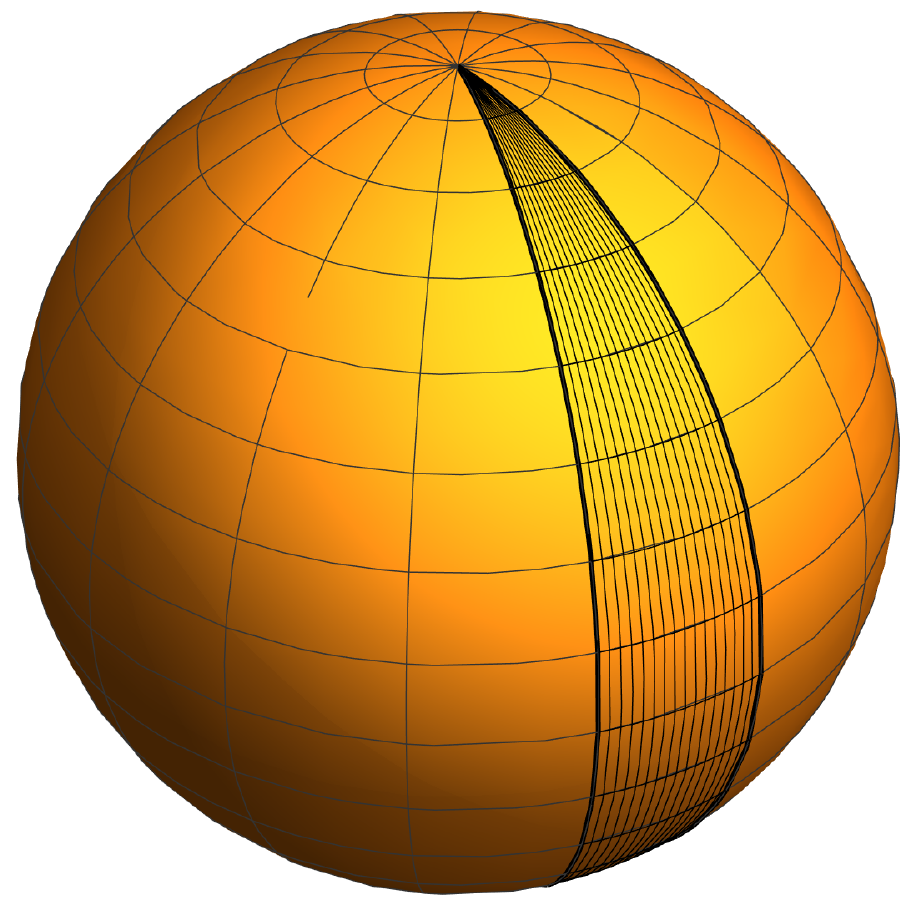}\caption{$S^2/\mathbb{Z}_n$}\label{fig:o1}
\end{subfigure}
\qquad \qquad
\begin{subfigure}[H]{0.4\textwidth}
\includegraphics[width=\textwidth]{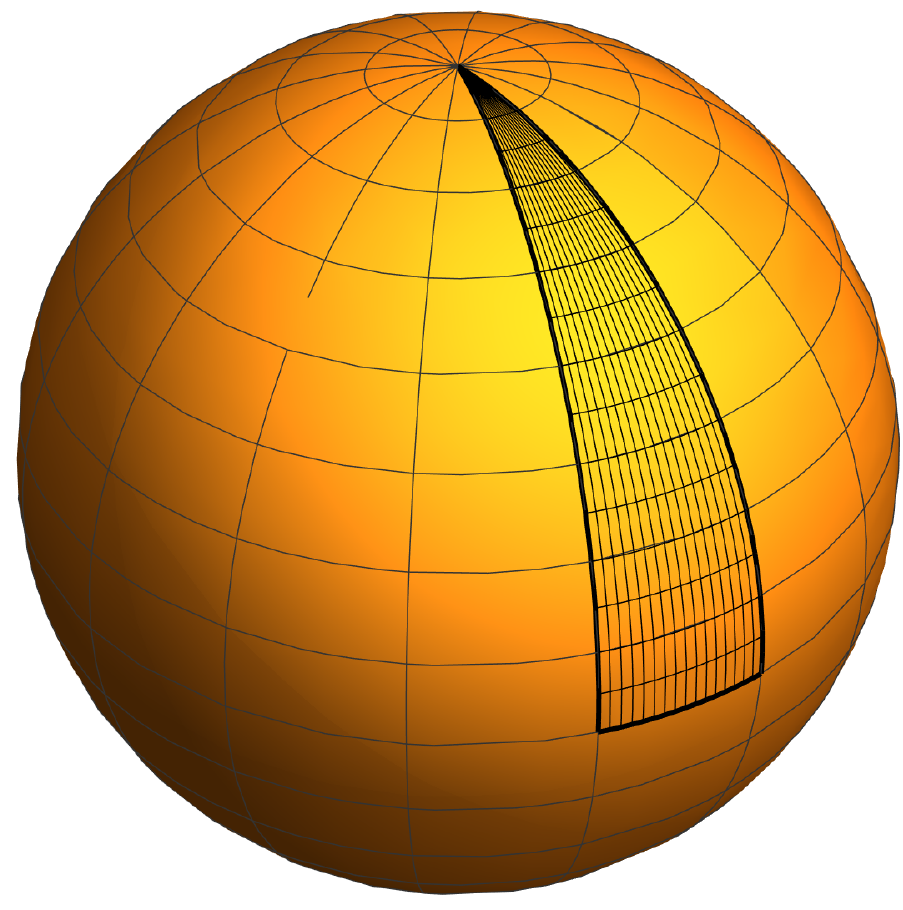}\caption{$S^2/D_n$}\label{fig:e1}
\end{subfigure}
\caption{The fundamental domains of the $S^2/\mathbb{Z}_n$  and $S^2/D_n$ orbifolds, depicted as the shaded regions of the sphere in \eqref{fig:o1} and \eqref{fig:e1} respectively.
}\label{fig:f1}
\end{center}
\end{figure}

The orbifolding thus decreases the size of the longitudinal direction by a factor of $n$, and the latitudinal direction by a factor of 2, as depicted in fig.~\eqref{fig:e1}. 
This is also reflected in the fact that the lowest non-trivial scalar Laplacian eigenvalue occurs at $k=2$, instead of $k=1$. To see this, 
 note that there are no harmonic polynomials at level $k=1$ which are invariant under the orbifold generators, \cf~\S\ref{app:a1}. 
 At level $k=2$,  on the other hand,  $p_2(\vec{x})=x^2+y^2-2z^2$ is harmonic and invariant under both orbifold generators. We thus obtain an order of scale separation $r=3$, \cf~\eqref{a8}.

\subsection{$S^2/\mathcal{T}$}
 
The $S^2/\mathcal{T}$ orbifold is generated by parity transformations in any two coordinates, and cyclic permutations of the coordinates. It can be seen that at levels $k=1,2$ no harmonic polynomial is invariant under the orbifold generators, while at level $k=3$ the polynomial $p_3(\vec{x})=xyz$ is harmonic and invariant. This results in an order of scale separation $r=6$.

\subsection{$S^2/\mathcal{O}$}

It can be seen that the octahedral group does not leave any harmonic polynomials invariant at levels $k=1,\dots,3$. At level $k=4$ there is a unique harmonic invariant, 
\eq{\label{2.35}
p_4(\vec{x})=x^4 + y^4 + z^4- 3 x^2 y^2 - 3 x^2 z^2 - 3 y^2 z^2
~.}
We thus have an order of scale separation $r=10$.

\subsection{$S^2/\mathcal{I}$}\label{sec:s2I}

In the icosahedral case, 
it can be seen that at levels $k=1,\dots,5$, no harmonic polynomial is invariant under the orbifold generators, \cf~\eqref{a5}.~At level $k=6$ there is a unique harmonic 
polynomial \cite{spherical_harmonics_computer} given by (up to overall normalization),
\eq{\label{2.3}
p_6(\vec{x})=
(x^2 + y^2)^3 + \frac{42}{5}  (x^4 - 10 x^2 y^2 + 5 y^4)x z - 
 18 (x^2 + y^2)^2 z^2 + 24 (x^2 + y^2) z^4 - \frac{16}{5} z^6
~.}
This results in an order of scale separation $r=21$.

\section{Some invariant theory}\label{sec:molien}

In the previous sections we have constructed invariant harmonics using brute force.~We will now introduce some more sophisticated machinery which simplifies the task at hand, 
issued from invariant theory, see e.g.~\cite{sturmfels}. In section \S\ref{sec:s2I2} we will apply this technology to reproduce the result for $S^2/\mathcal{I}$ from \S\ref{sec:s2I}.

Starting from any polynomial function $f(\vec{x})$, an invariant $I_f$ (which may be identically zero) can be obtained through the so-called 
Reynolds operator,
\eq{\label{rey}
I_f(\vec{x})= \frac{1}{|R(G)|}\sum_{g\in R(G)}{f(g\circ\vec{x})}
~,}
where $|R(G)|$ denotes the number of elements in $R(G)$.\footnote{The distinction between $|R(G)|$ and the  the order of $G$ is 
relevant only if $R(G)$ is not faithful.}~For a finite matrix group $G$ and a representation $R(G)$ thereof, the number of   invariant 
homogeneous polynomials (not necessarily harmonic) in that representation is generated by the Molien function \cite{meyer,Patera:1978qx}, given by,
\eq{\label{mol}
M_{R(G)}(\lambda)\equiv\frac{1}{|R(G)|}\sum_{g\in R(G)}\frac{1}{\det(\mathbb{I}-\lambda g)}
~.}
More specifically, the number $g_m$ of {\it linearly}-independent invariant 
polynomials of degree $m$ is the coefficient of $\lambda^m$ in the expansion of the Molien 
function, 
\eq{
M_{R(G)}(\lambda)=\sum_{m=0}^\infty g_m\lambda^m
~.
}
The subring of polynomials invariant under the action of $R(G)$ is generated by  a minimal set of so-called fundamental invariants. This means that all  $R(G)$-invariants can be expressed as polynomials of the fundamental invariants.  
The latter are further divided into a set of primary invariants $\{I_{m_1},\dots,I_{m_r}\}$ of degrees ${m_1},\dots, {m_r}$, which are {\it algebraically}-independent,\footnote{
The polynomials $p_1, \dots, p_N$ are called algebraically-dependent if there exists a (not identically vanishing) polynomial in $N$ variables $h(x_1,\dots x_N)$, such that  
$h(p_1,\dots p_N)=0$.
} and a set of secondary invariants $\{1,\bar{I}_{n_1},\dots,\bar{I}_{n_s}\}$ of degrees $0,{n_1},\dots, {n_s}$, which are not.~The number $r$ of primary invariants is equal to the dimension of the representation $R(G)$ \cite{burnside},\footnote{The 
algebraic independence of the primary invariants is straightforward to check using the Jacobian criterion \cite{humphreys_1990}. The latter  states that 
the $N$ polynomials  $p_1, \dots, p_N$ in $N$ variables $x_1,\dots,x_N$ are   algebraically independent if and only if,
\eq{\det\left(
\frac{\partial p_i}{\partial x_j}
\right)\neq 0
~.}
}
 while the number of secondary invariants is equal to  \cite{sturmfels}, 
\eq{\label{3.3}
1+s=\frac{m_1\dots{m_r}}{|G|}
~.}
The algebraic relations between primary and secondary invariants  are called 
syzygies, and are of the form,
\eq{
\bar{I}_{n_i}^2=f(I_{m_1},\dots,I_{m_r})+\sum_j\bar{I}_{n_j}g_j(I_{m_1},\dots,I_{m_r})
~,}
for some polynomial functions $f$, $g_j$ of the primary invariants. 
The Molien function can be put in the form,
\eq{
M_{R(G)}(\lambda)=\frac{1+\sum_{n}a_{n}\lambda^{n}}{(1-\lambda^{m_1})\dots(1-\lambda^{m_r})}
~,}
where $a_{n}$ is the number of secondary invariants of degree $n>0$. In particular, $\sum_{n}a_{n}=s$, \cf~\eqref{3.3}.

Let us now consider the case where $G$ is a finite group of orthogonal matrices acting on the variables $\vec{x}$ in the fundamental representation.~A straightforward generalization of the proof of \cite{meyer} gives the generating function of {\it harmonic} invariants, 
\eq{\label{hgen}
\mathfrak{h}(\lambda)=(1-\lambda^2)M_{R(G)}(\lambda)
~.}

\subsection{$S^2/\mathcal{I}$ again}\label{sec:s2I2}

Let us return to the case of the icosahedral group and reproduce the results of \S\ref{sec:s2I} using the machinery of \S\ref{sec:molien}.~The group $\mathcal{I}$ acts on the coordinates $(x,y,z)$ of $\mathbb{R}^3$ in the three-dimensional representation. 
Computing the Molien function \eqref{mol} 
we obtain,\footnote{We have used a simple Mathematica code to generate all group elements, and compute the 
Reynolds operator and the Molien function in a series expansion.  For some powerful publicly available 
software see {\it GAP} \cite{gap-system} and {\it SUTree} \cite{Merle:2011vy} }
\eq{\label{3.7}
M_{\bf{3}}(\lambda)=\frac{1+ \lambda^{15}}{(1-\lambda^{2})(1-\lambda^{6})(1-\lambda^{10})}
=1 + \lambda^2 + \lambda^4 + 2 \lambda^6+\dots
~.}
This suggests the existence of three primary invariants of degrees two, six and ten and one secondary invariant of degree 15, as can be verified 
by constructing the syzygies \cite{Merle:2011vy}. 
In addition there is one invariant of degree four and one of degree six which are not independent.

Using the explicit matrix  representation of $\mathcal{I}$ given in \eqref{explicitrlrmrnts}, 
the degree-2 invariant can be constructed by acting with  the Reynolds operator \eqref{rey} on the monomial $f=x^2$,
\eq{I_{x^2}=\frac13(x^2+y^2+z^2)
~.}
This invariant simply reflects the fact that $\mathcal{I}$ is a subgroup of $SO(3)$.~Similarly, starting from $f=x^4$ we obtain the invariant of degree four,
\eq{I_{x^4}=\frac15(x^2+y^2+z^2)^2
~,}
which, as expected from the discussion below \eqref{3.7},  is not algebraically independent as it is proportional to $I_{x^2}^2$. At degree six we can construct two invariants 
acting with  the Reynolds operator on  $f=x^6$ and $f=z^6$, respectively,
\eq{\spl{\label{3.11}
240I_{x^6}&=35 (x^2 + y^2)^3 + 6 x (x^4 - 10 x^2 y^2 + 5 y^4) z + 
 90 (x^2 + y^2)^2 z^2 + 120 (x^2 + y^2) z^4 + 32 z^6\\
 75I_{z^6}&=10 (x^2 + y^2)^3 - 6 x (x^4 - 10 x^2 y^2 + 5 y^4) z + 
 45 (x^2 + y^2)^2 z^2 + 15 (x^2 + y^2) z^4 + 13 z^6
~.}}
As expected these are not algebraically independent, but satisfy,
\eq{
16 I_{x^6} + 5 I_{z^6} - 45 I_{x^2} I_{x^4}=0
~.}
It is now easy to verify that neither $I_{x^2}$, nor $I_{x^4}$ are harmonic, in agreement with  the result of \S\ref{sec:s2I}.
Moreover,  the only harmonic linear combination (up to overall normalization) of degree six is given by $I_{x^6} - I_{z^6}$, 
which is proportional to  the  polynomial already determined in \eqref{2.3}. This is of course consistent with \eqref{hgen}, \eqref{3.7}, 
\eq{
\mathfrak{h}(\lambda)= 1+\lambda^6+\dots
~,}
which indicates that the first non-trivial  harmonic invariant appears at degree six.

\section{Orbifolds of $S^5$}\label{sec:4}

The  orbifolds of $S^5$ are obtained by orbifolding by a finite subgroup $\Gamma\subset SO(6)$, viewed as acting on the coordinates of the ambient $\mathbb{R}^6$.~Since it is expected that only quotients for which $\Gamma\subset SU(3)$ lead to a  supersymmetric theory \cite{Douglas:1997de},  we will focus on this case.~The group  
$\Gamma$ acts on the coordinates $(x,y,z)$ of the ambient $\mathbb{C}^3\simeq \mathbb{R}^6$ in the fundamental, three-dimensional representation {\bf 3}.~We are now interested in invariants 
which are (complex) polynomials in the six variables $(x,y,z,x^*,y^*,z^*)$,\footnote{Real polynomials can be constructed by taking real and imaginary parts thereof.}  on which $\Gamma$ acts block-diagonally in the ${\bf 3}\oplus{\bf \bar{3}}$ representation. 
The relevant Molien function is,
\eq{\label{mol6}
M_{{\bf 3}\oplus{\bf \bar{3}}}(\lambda) =\frac{1}{|\Gamma|}\sum_{g\in\Gamma}\frac{1}{\det(\mathbb{I}-\lambda g)\det(\mathbb{I}-\lambda g^*)}
~,}
where $|\Gamma|$   is the order of $\Gamma$. On the other hand, the Molien function, 
\eq{\label{mol6h}
M_{{\bf 3}}(\lambda) =\frac{1}{|\Gamma|}\sum_{g\in\Gamma}\frac{1}{\det(\mathbb{I}-\lambda g)}
~,}
enumerates the holomorphic invariants of $\Gamma$ (and of course also the antiholomorphic ones, by complex conjugation). We will therefore refer to it as the ``holomorphic Molien function''. Note that the holomorphic and antiholomorphic invariants are automatically harmonic with respect to the Laplacian on $\mathbb{R}^6$,
\eq{
\Delta=\frac{\partial^2}{\partial x\partial x^*}+\frac{\partial^2}{\partial y\partial y^*}+\frac{\partial^2}{\partial z\partial z^*}
~.}
A classification of finite subgroups of $SU(3)$ appeared over a century ago in \cite{MillerBlichfeldtDickson}.~A more detailed analysis,  involving character tables and generators,  motivated by particle physics applications appeared in \cite{FairbairnFultonKlink}. Further additions to the list of  \cite{FairbairnFultonKlink} appeared subsequently in \cite{BovierLulingWyler1,BovierLulingWyler2,FairbairnFulton}. 
A comprehensive review of the subject can be found in \cite{Ludl:2009ft,Ludl:2011gn}.

Discrete subgroups of $SU(3)$ have been considered in particle physics model-building, going back to the work of \cite{FairbairnFultonKlink}.   
More recently they have been used in flavor physics applications,  
as a means to constrain Yukawa couplings, mass matrices and mixing angles  in the quark and lepton sectors (for reviews and further references see {\it e.g.} \cite{Altarelli:2010gt, Ishimori:2010au}).

The classification of finite  subgroups of $SU(3)$ consists of:\footnote{We follow the presentation of \cite{Grimus:2011fk} which includes some minor corrections to 
the list of \cite{MillerBlichfeldtDickson,FairbairnFultonKlink}. }
\begin{itemize}

\item Abelian groups of diagonal matrices. These are isomorphic to $\mathbb{Z}_m\times \mathbb{Z}_n$, where $n$ is a divisor of $m$. 
\item Finite subgroups of $U(2)$ of the form,
\eq{\label{u2}
\left( \begin{array}{cc}
 \det A^* &0  \\
  0 & A
  \end{array} \right)~;~~~A\in U(2)
~.}
\item The $C(n,a,b)$ series generated by,
\eq{\label{EF}
E=
\left( \begin{array}{ccc}
 0&1 &0  \\
  0 & 0& 1 \\
  1&0&0
  \end{array} \right)~;~~~
  F(n,a,b)=
  \left( \begin{array}{ccc}
 \eta^a&0 &0  \\
  0 & \eta^b& 0 \\
  0&0 & \eta^{-a-b}
  \end{array} \hskip-.1cm\right)
~,}
where $n\in\mathbb{N}-\{0\}$; $\eta\equiv e^{2\pi i/n}$; $a,b\in\mathbb{N}$ with $a,b\leq n-1$.
\item The $D(n,a,b;d,r,s)$ series generated by $E$, $F(n,a,b)$ given above and,
\eq{\label{G}
G(d,r,s)=
\left( \begin{array}{ccc}
 \delta^r&0 &0  \\
  0 & 0& \delta^s \\
  0&-\delta^{-r-s}&0
  \end{array} \right)~,
 }
where $d\in\mathbb{N}-\{0\}$; $\delta\equiv e^{2\pi i/d}$; $r,s\in\mathbb{N}$ with $r,s\leq d-1$.
\item The ``exceptional'', or ``crystallographic''  groups,\footnote{We consider $\varphi=3$, in the notation of \cite{FairbairnFultonKlink}. The case $\varphi=1$ corresponds to subgroups 
of $SU(3)/\mathbb{Z}_3$.} 
\eq{\Sigma(60)~,~  \Sigma(60)\times \mathbb{Z}_3~, ~\Sigma(168)~,  \Sigma(168)\times \mathbb{Z}_3~,~ \Sigma(36\varphi)~,  
 ~\Sigma(72\varphi)~,   ~\Sigma(216\varphi)~,   ~\Sigma(360\varphi)~,
}
whose explicit generators can be found in {\it e.g.} \cite{Ludl:2009ft}.
\end{itemize}
The relation of the well-known 
$\Delta$ and $T_n$ series of $SU(3)$ subgroups to the above classification is as follows.~The  $T_n$ subgroups are special cases of the $C$-series.~On the other hand, 
it was shown in \cite{Ludl:2009ft} that 
all $SU(3)$ subgroups in the  $C$-series can be interpreted as three-dimensional irreducible
representations of $\Delta(3n^2)$,  which were determined in \cite{Luhn:2007uq}. Moreover, 
  \cite{Zwicky:2009vt} showed that every $SU(3)$ subgroup in the  $D$-series   is a subgroup of some $\Delta(6n^2)$.\footnote{The subgroups in the $D$-series cannot all  be interpreted 
as irreducible three-dimensional representations of $\Delta(6n^2)$ \cite{Ludl:2011gn}.} More explicitly, we have the following subgroup structure (see   \cite{Merle:2011vy} for the complete finite $SU(3)$-subgroup tree),
\eq{
C(n,a,b)\subseteq \Delta(3n^2)~;~~~
D(m,a,b;d,r,s) \subseteq \Delta(6n^2)
~,}
where in the second relation above $n$ is equal to the lowest common multiple of 2, $m$, $d$. In addition,\footnote{Note, however, that $T_n$ is not defined for all $n\in\mathbb{N}$, {\it cf.}~\S\ref{sec:tn}.} 
\eq{
T_n\subset \Delta(3n^2)  \subset \Delta(6n^2)
~,}
where both subgroup relations are maximal.~Moreover, within the exceptional series, the following subgroup structure holds,
\eq{\Sigma(36\varphi)\subset \Sigma(72\varphi)\subset \Sigma(216\varphi)~;~~~\Sigma(60)\subset \Sigma(60)\times \mathbb{Z}_3\subset \Sigma(360\varphi)
~.}
Consider the case  $H\subset G\subset SU(3)$.~Clearly all $G$-invariants are also $H$-invariants.~It follows that the $G$-invariant KK spectrum is a subset of the $H$-invariant KK spectrum.~In particular,  $m_{\text{KK}/G}\geq m_{\text{KK}/H}$,  hence modding out the internal space with $G$  leads, in general, to  higher scale separation than modding out with  $H$, {\it cf.}~\eqref{1}.~Therefore the highest scale separation is achieved for maximal   finite subgroups of $SU(3)$.\footnote{A subgroup $G$ is called maximal if there is no 
other proper subgroup $G'$ such that  $G\subset G'\subset SU(3)$. }

\subsection{$S^5/\mathbb{Z}_m\times\mathbb{Z}_n$}\label{sec:s5Iz}

These are abelian groups of diagonal matrices. Their elements are of the general form,\footnote{Let $m$ be the maximal order of an element of the abelian group.~Defining $\mu\equiv e^{2\pi i/m}$, it can be seen that every element of the group can be put in the form of eq.~\eqref{ccc} with $a=\mu^c$, $b=\mu^d$,  and $c,d\in\{0,\dots,m-1\}$. 
It follows ({\it cf}.~Theorem 2.1 and \S A of \cite{Ludl:2011gn}) that the resulting group must be isomorphic to $\mathbb{Z}_m\times\mathbb{Z}_n$ with $n$ a divisor of $m$.}
\eq{\label{ccc}
\left( \begin{array}{ccc}
 a\hskip .2cm{} &0\hskip .1cm{}  &0  \\
  0\hskip .2cm{}  & b\hskip .1cm{} &0 \\
  0\hskip .2cm{} &0&\hskip .1cm{} a^*b^*
  \end{array} \right)~,~~~a,b\in U(1)
~,}
so they manifestly leave invariant the quadratic 
harmonic polynomials, $|x|^2-|z|^2$, $|y|^2-|z|^2$, which corresponds to an order of scale separation $r=2.4$.

\subsection{$S^5/\Gamma_{U(2)}$}\label{sec:s5Iu}

The quadratic 
harmonic  polynomial, $|x|^2+|y|^2-2|z|^2$, is invariant under the action of elements of the general form \eqref{u2}, thus resulting in an order of scale separation $r=2.4$.

\subsection{$S^5/\Sigma(60)$}\label{sec:s5I}

The group $\Sigma(60)$ turns out to be isomorphic to the icosahedral group: $\Sigma(60)\simeq \mathcal{I}\subset SO(3)$, already considered in section \S\ref{sec:2}. 
Computing the Molien function \eqref{mol6} we obtain,
\eq{\label{411m}
M_{{\bf 3}\oplus{\bf {3}}}(\lambda) =1+3\lambda^2+6\lambda^4+17\lambda^6+\dots
~,}
where we have noted that the fundamental representation of $\Sigma(60)$ is real.~Three linearly-independent quadratic invariants are,
\eq{\label{410}
I_{2,1}= x^2+y^2+z^2~;~~~I_{2,2}=I_{2,1}^*~;~~~I_{2,3}=|x|^2+|y|^2+|z|^2
~,}
constructed by acting with  the Reynolds operator on $x^2$, $x^{\star2}$, $|x|^2$ respectively.~The first two are $SO(3)$ invariants, while the third one is an $SU(3)$ invariant.~More generally, for all $n\in\mathbb{N}$, there is exactly one invariant of $SU(3)$ at order $2n$, proportional to $I_{2,3}^n$. This invariant is never harmonic, except for $n=0$. 
On the other hand, $I_{2,1}$, $I_{2,2}$ are harmonic (since they are holomorphic and antiholomorphic respectively). 
We thus obtain an order of scale separation $r=2.4$.

Six linearly-independent quartic invariants are,
\eq{\label{411}
I_{2,1}^2~;~~~I_{2,2}^2;~~~I_{2,3}^2~;~~~I_{2,1}I_{2,2}~;~~~I_{2,1}I_{2,3};~~~I_{2,2}I_{2,3}
~.}
At degree six we have 17 linearly-independent sextic polynomials, including the $I_{x^6}$, $I_{z^6}$ of \eqref{3.11} and their complex conjugates. 
The latter four invariants are all harmonic.

\subsection{$S^5/\Sigma(60)\times \mathbb{Z}_3$}\label{sec:s5I2}

The group $\Sigma(60)\times \mathbb{Z}_3$ is generated by $A\equiv e^{2\pi i/3} \mathbb{I}$, together with the generators of $\Sigma(60)$. 
Therefore the invariants of $\Sigma(60)\times \mathbb{Z}_3$ are the subset of the invariants of $\Sigma(60)$ which are also invariant under $A$. Since 
$A$ multiplies $(x,y,z)$ by  $e^{2\pi i/3}$, 
clearly  $I_{2,1}$, $I_{2,2}$  of \eqref{410} are not invariant under $A$.  On the other hand $I_{2,3}$ is invariant under $A$, however it is not harmonic. 

The only quartic invariants of $\Sigma(60)$ which are also invariant under  $A$, are $I_{2,1}I_{2,2}$, $I_{2,3}^2$, {\it cf.}~\eqref{411}, neither of which is harmonic. However their linear combination $I_{2,3}^2-2I_{2,1}I_{2,2}$ is indeed harmonic. We thus obtain an order of scale separation $r=6.4$.

At degree six, the harmonic invariants $I_{x^6}$, $I_{z^6}$  and their complex conjugates, already discussed 
in \S\ref{sec:s5I}, are also invariant under $A$.

\subsection{$S^5/\Sigma(168)$}\label{sec:s5Iq}

The group $\Sigma(168)$ is generated by,
\eq{
 S'= \frac{i}{\sqrt{7}}
\left( \begin{array}{ccc}
 \eta^2-\eta^5& \eta-\eta^6 & \eta^4-\eta^3  \\
   \eta-\eta^6 &  \eta^4-\eta^3& \eta^2-\eta^5 \\
   \eta^4-\eta^3& \eta^2-\eta^5& \eta-\eta^6
  \end{array} \right)
~;~~~
 T'=\frac{i}{\sqrt{7}}
\left( \begin{array}{ccc}
 \eta^3-\eta^6& \eta^3-\eta & \eta-1  \\
   \eta^2-1 &  \eta^6-\eta^5& \eta^6-\eta^2 \\
   \eta^5-\eta^4& \eta^4-1& \eta^5-\eta^3
  \end{array} \right)
   ~,}
where $\eta\equiv e^{2\pi i/7}$. 
Computing the Molien function \eqref{mol6} we obtain,
\eq{\label{411mq}
M_{{\bf 3}\oplus{\bf {\bar{3}}}}(\lambda) =1+\lambda^2+3\lambda^4+8\lambda^6+\dots
~.}
The quadratic invariant is $I_{2,3}$ of \eqref{410}. Three linearly-independent quartic invariants are $I_{2,3}^2$, together with,
\eq{\label{410q}
I_{4,1}=x y^3 + x^3 z + y z^3
~,}
and its complex conjugate. Since  $I_{4,1}$ is holomorphic, it is also harmonic.~We thus obtain an order of scale separation $r = 6.4$.

\subsection{$S^5/\Sigma(168)\times \mathbb{Z}_3$}\label{sec:s5Iqp}

The group $\Sigma(168)\times \mathbb{Z}_3$  is generated by $A\equiv e^{2\pi i/3} \mathbb{I}$, together with the generators of $\Sigma(168)$. 
Therefore the invariants of $\Sigma(168)\times \mathbb{Z}_3$ are the subset of the invariants of $\Sigma(168)$ which are also invariant under $A$. 
At quadratic and quartic level, only $I_{2,3}$ and $I_{2,3}^2$ respectively are invariant under $A$, {\it cf.}~\S\ref{sec:s5Iq}.~However neither of these is harmonic.~Hence 
$S^5/\Sigma(168)\times \mathbb{Z}_3$ does not have any harmonic invariants of degree less than five.~Moreover,  $\Sigma(168)\times \mathbb{Z}_3$ 
does not have any invariants of degree five either, since its subgroup $\Sigma(168)$ does not, as follows from the Molien function \eqref{411mq}.   
 At degree six, the following polynomial, obtained by acting with the Reynolds operator on $(xyz)^2$,  is invariant under the group generators,
 \eq{
 I_{6,1}=x^5 y + y^5 z + z^5x- 5 x^2 y^2 z^2 
 ~.}
Since  $I_{6,1}$ is holomorphic, it is also harmonic. Hence the order of scale separation is $r=12$.

\subsection{$S^5/\Sigma(36\varphi)$}\label{sec:s532}

The group $\Sigma(36\varphi)$ is be generated by,
\eq{
 A= 
\left( \begin{array}{ccc}
 1&0 &0  \\
  0 & \omega&0 \\
  0&0&\omega^2
  \end{array} \right)
~;~~~
  B= 
\left( \begin{array}{ccc}
 0&1 &0  \\
  0 & 0& 1 \\
  1&0&0
  \end{array} \right)
  ~;~~~
 C=\frac{1}{\omega-\omega^2}
\left( \begin{array}{ccc}
 1&1 &1  \\
  1 & \omega& \omega^2 \\
  1&\omega^2&\omega
  \end{array} \right)
   ~,}
where $\omega\equiv e^{2\pi i/3}$. 
Computing the Molien function \eqref{mol6} we obtain,
\eq{\label{415m}
M_{{\bf 3}\oplus{\bf \bar{3}}}(\lambda) =1+\lambda^2+2\lambda^4+2\lambda^5+8\lambda^6+\dots
~.}
At quadratic and quartic level, we have the $I_{2,3}$ and $I_{2,3}^2$ invariants respectively.~There is one additional quartic  invariant, 
\eq{I_{4,2}= 
(x^2-yz)~\!y^*z^*+(y^2-xz)~\!x^*z^*+(z^2-xy)~\! x^*y^*
 + \text{c.c.}}
The combination $2I_{4,2}+I_{2,3}^2$ is  harmonic, so  we  obtain an order of scale separation $r=6.4$.

\subsection{$S^5/\Sigma(72\varphi)$}\label{sec:s5I2.5}

The group $\Sigma(72\varphi)$ is obtained by combining  $\Sigma(36\varphi)$ with the generator, 
\eq{
D=\frac{1}{\omega-\omega^2}
\left( \begin{array}{ccc}
 1&1 &\omega^2  \\
  1 & \omega& \omega \\
  \omega & 1&\omega
  \end{array} \right)
   ~.}
Computing the Molien function \eqref{mol6} we obtain,
\eq{\label{415m}
M_{{\bf 3}\oplus{\bf \bar{3}}}(\lambda) =1+\lambda^2+\lambda^4+2\lambda^5+4\lambda^6+ \dots
~.}
At quadratic and quartic level, we have the $I_{2,3}$ and $I_{2,3}^2$ invariants respectively, which are not harmonic. 
Two quintic linearly-independent invariants can be constructed,
\eq{
\spl{
I_{5,1}=|x|^2(y^3 - z^3) + |y|^2 (z^3  - x^3)  + |z|^2 ( x^3- y^3 )~;~~~
I_{5,2}= I_{5,1}^*
~,}}
by acting with the Reynolds operator on $|x|^2 y^3$, $|x|^2 y^{*3}$, respectively. Both of these invariants are harmonic, 
so we obtain an order of scale separation $r=9$.

\subsection{$S^5/\Sigma(216\varphi)$}\label{sec:s5I2.5}

The group $\Sigma(216\varphi)$ can be generated by the following two generators \cite{Ludl:2009ft},
\eq{
 A=\frac{1}{\omega-\omega^2}
\left( \begin{array}{ccc}
 1&1 &1  \\
  1 & \omega& \omega^2 \\
  1&\omega^2&\omega
  \end{array} \right)
~;~~~
  B=\epsilon
\left( \begin{array}{ccc}
 1&0 &0  \\
  0 & 1& 0 \\
  0&0&\omega
  \end{array} \right)
   ~,}
where $\epsilon\equiv e^{4\pi i/9}$, $\omega\equiv e^{2\pi i/3}$. 
Computing the Molien function \eqref{mol6} we obtain,
\eq{\label{415m}
M_{{\bf 3}\oplus{\bf \bar{3}}}(\lambda) =1+\lambda^2+\lambda^4+2\lambda^6+3\lambda^8+4\lambda^9+\dots
~.}
At quadratic, quartic and sextic order, we have the $I_{2,3}$, $I_{2,3}^2$ and $I_{2,3}^3$ invariants respectively, which are not harmonic. 
At order six, there is one additional invariant, 
\eq{
I_{6,2}=18|xyz|^2+(x^3+y^3+z^3)(x^{\star3}+y^{\star3}+z^{\star3})
~,}
which can be constructed by acting with  the Reynolds operator on $|xyz|^2$. 
This invariant is not harmonic either, however it can be seen that the linear combination $5I_{6,2}-3I_{2,3}^3$ is indeed harmonic. We thus obtain an order of scale separation $r=12$. 

\subsection{$S^5/\Sigma(360\varphi)$}\label{sec:s5I3}

The generators of $\Sigma(360\varphi)$ can be taken to be,
\eq{\label{415g}
A=
\left( \begin{array}{ccc}
 0&1 &0  \\
  0 & 0& 1 \\
  1&0&0
  \end{array} \right)~;~~~
  B=
\left( \begin{array}{ccc}
 1&0 &0  \\
  0 & -1& 0 \\
  0&0&-1
  \end{array} \right)~;~~~
  C=\frac12
\left( \begin{array}{ccc}
 -1&\mu_2 &\mu_1  \\
  \mu_2 & \mu_1& -1 \\
  \mu_1&-1&\mu_2
  \end{array} \right)~;~~~
  D=
\left( \begin{array}{ccc}
 -1&0 &0  \\
  0 & 0& -\omega \\
  0&-\omega^2&0
  \end{array} \right)  
~,}
where $\mu_1\equiv\tfrac12 (-1+\sqrt{5})$,  $\mu_2\equiv-\tfrac12 (1+\sqrt{5})$, $\omega\equiv e^{2\pi i/3}$. On the other hand $\Sigma(60)$ is generated 
by $A$, $B$ and $C$, so that $\Sigma(60)\subset \Sigma(360\varphi)$. Moreover, 
\eq{(A\cdot D)^2=\omega\hskip 0.05cm \mathbb{I}
~,}
which generates $\mathbb{Z}_3$. It follows  that  $\Sigma(60)\times \mathbb{Z}_3\subset \Sigma(360\varphi)$, and therefore  all $\Sigma(360\varphi)$ invariants 
are also $\Sigma(60)\times \mathbb{Z}_3$ invariants. From \S\ref{sec:s5I2} we then conclude that  $\Sigma(360\varphi)$ does not have any harmonic invariants of degree less than four. To check whether or not  $\Sigma(360\varphi)$ has a degree-four harmonic invariant, it suffices to check whether $I_{2,3}^2-2I_{2,1}I_{2,2}$ of \S\ref{sec:s5I2} is invariant under the generator $D$ of \eqref{415g}. It can be seen that it is not, therefore  $\Sigma(360\varphi)$ does not have any harmonic invariants of degree less than five. Moreover,  $\Sigma(360\varphi)$ does not have any invariants of degree five either, since its subgroup $\Sigma(60\varphi)$ does not, as follows from the Molien function \eqref{411m}. 
On the other hand, $\Sigma(360\varphi)$ has one holomorphic (and therefore harmonic) invariant of degree six, constructed explicitly in \cite{Merle:2011vy}. 
We thus obtain an order of scale separation $r=12$, {\it cf}.~\eqref{a8}.

\subsection{$T_n$}\label{sec:tn}

 The series $T_n$ is generated by $E$ and $F(n,1,b)$ of \eqref{EF}, with $1+b+b^2\equiv  0~\! (\text{mod}~\!n)$,  $n\geq2$.  The smallest group in this series is obtained for $n=7$, $b=2$. 
All  $T_n$'s admit a cubic  
harmonic invariant, $x yz$, which corresponds to an order of scale separation $r=4.2$.

\subsection{$\Delta(3n^2)$}  
This series is generated by $E$ and $F(n,0,1)$ of \eqref{EF}, with $n\geq 2$. 
It   admits a cubic  
harmonic invariant, $x yz$, which corresponds to an order of scale separation $r\leq 4.2$.  For  $n=2$ we have $\Delta(12)\simeq A_4$,  and  the additional quadratic harmonic invariants $I_{1,2}$ of \eqref{410}, 
which gives an order of scale separation $r=2.4$.

 \subsection{$\Delta(6n^2)$}

This series is generated by $E$,   $F(n,0,1)$ and $G(2,1,1)$ of \eqref{EF}, \eqref{G}, with $n\geq 2$.  It does not admit any cubic harmonic invariants, but  
it admits  the quartic harmonic invariant, $2(|x|^4+|y|^4+|z|^4)-I_{2,3}^2$, which corresponds to an order of scale separation $r\leq 6.4$. For  $n=2$ we have $\Delta(24)\simeq S_4$, and  the additional quadratic harmonic invariants $I_{2,1}$, $I_{2,2}$ of \eqref{410}, 
which corresponds to an order of scale separation $r=2.4$.

\section{Smoothness and Killing spinors}\label{sec:smooth}

For the background AdS$_p\times S^q/\Gamma\times M_d$ to preserve supersymmetry, we would typically need the existence of Killing spinors on the orbifolded sphere.~On the other hand, 
it is well known that Killing spinors on $S^q$ are in one-to-one correspondence with (covariantly) constant spinors on the ambient $\mathbb{R}^{q+1}$, so that each Killing spinor   is the restriction 
to $S^q$ of a constant spinor of $\mathbb{R}^{q+1}$ \cite{Bar:1993gpi}. Therefore, 
if $\tilde{\Gamma}$ is a lift of $\Gamma$ to Spin$(q+1)$ corresponding to the spin structure of $S^q/\Gamma$, Killing spinors of the latter would correspond to $\tilde{\Gamma}$-invariant spinors.  
In particular, for $q=5$, 
taking $\Gamma\subset SU(3)\subset SU(4)\simeq$~Spin(6) thus guarantees the existence of Killing spinors on $S^5/\Gamma$. 

The subgroups $\Gamma\in SO(6)$ for which $S^5/\Gamma$ is smooth,  
were classified in \cite{wolf}. Moreover, there is an infinite number of smooth quotients possessing Killing spinors \cite{sulanke,Acharya:1998db}. 
Rephrasing the results presented in \cite{Acharya:1998db}, by appropriately choosing  the complex structure of the ambient $\mathbb{C}^3$, there are two infinite series of smooth $S^5/\Gamma$ orbifolds 
possessing Killing spinors, 
\begin{itemize}
\item $\Gamma$  isomorphic to $\mathbb{Z}_n$, generated by, 
\eq{\label{s1}
\left( \begin{array}{ccc}
 \eta&0 &0  \\
  0 & \eta^a& 0 \\
  0&0&\eta^b
  \end{array} \right)
~,}
where $\eta\equiv e^{2\pi i/n}$, $n\in\mathbb{N}-\{0\}$; 
$a,b\in\mathbb{Z}$ with $a+b+1\equiv  0~\! (\text{mod}~\!n)$, $1\leq |a|,|b|\leq n-1$, and $(a,n)=(b,n)=1$, where  $(p,q)$ denotes the greatest common divisor of $p$, $q$.

\item $\Gamma$  isomorphic to $T_n$, generated by $E$ and $F(n,1,b)$ of \eqref{EF}, with $1+b+b^2\equiv  0~\! (\text{mod}~\!n)$. In addition we must impose $(3(b-1),n)=1$. 
These conditions only admit solutions for very specific integers $n$.\footnote{The general prime factor decomposition of such $n$ was determined in \cite{Grimus:2013apa}. In particular, 
if 3 is not a divisor of  $n$, as is the case here, $n$ must be a product of prime numbers, each of which is of the form $6k+1$, for positive integers $k$, and $b\notequiv 1~\! (\text{mod}~\!3)$. 
Ref.~\cite{Acharya:1998db} also lists  the conditions: $n$ odd, and $b\notequiv b^3\equiv  1~\! (\text{mod}~\!n)$. These  follow from the conditions already imposed in the main text.}
\end{itemize}
Both of these series are of the form $\Gamma\subset SU(3)$, which of course was expected in view of what was mentioned at the beginning of this section. The first of the two series admits quadratic 
harmonic invariants, linear combinations of $|x|^2-|z|^2$ and $|y|^2-|z|^2$, which corresponds to an order of scale separation $r=2.4$.~The second series admits a cubic  
harmonic invariant, $x yz$, which corresponds to an order of scale separation $r=4.2$.

Let us now come to the $S^2$ orbifolds. 
None of the these obifolds is smooth: their singularity types  are well understood and have been classified, see {\it e.g.}~Ch.13 of \cite{Thurston_orbifolds}. Moreover, $S^2/\Gamma$ does not 
admit Killing spinors, since there are no $SU(2)$-invariant spinors for any of the  subgroups  $\tilde{\Gamma}\subset SU(2)$, as can be easily verified using the generators  listed in \S\ref{app:su2}.

\section*{Acknowledgment}

I would like to thank Alexander Merle and Roman Zwicky for sharing with me  their {\it SUTree} files.

 \begin{appendix}

\section{Conventions}\label{app:conv}

The $p$-dimensional AdS  space can be defined as a the hyperboloid,
\eq{
x_0^2+x_p^2-\sum_{i=1}^{p-1}x_{i}^2=L_{\text{AdS}}^2~,
}
in an ambient $\mathbb{R}^{2,p-1}$ space, with a standard flat metric, parameterized by $x_0,\dots x_{p}$. 
The constant $L_{\text{AdS}}$ is the {\it radius of curvature} of AdS. 
The induced AdS$_p$ 
metric ($g_{\mu\nu}$) obeys,
\eq{
R_{\mu\nu}=-\frac{p-1}{L_{\text{AdS}}^2}~\!g_{\mu\nu}~, 
}
where $R_{\mu\nu}$ is the Ricci tensor of AdS$_p$. In local coordinates, covering half the hyperboloid, 
the metric takes the form,
\eq{
\d s^2(\text{AdS}_p)=L_{\text{AdS}}^2\left[
\d\rho^2 +e^{2\rho}\d s^2(\mathbb{R}^{1,p-2})
\right]
~.}
Similarly, the round $q$-dimensional sphere $S^q$ of radius $L_S$ is defined by, 
\eq{
 \sum_{i=1}^{q+1}x_{i}^2=L_{S}^2~,
}
in an ambient $\mathbb{R}^{q+1}$ space, with a standard flat metric, parameterized by $x_1,\dots x_{q+1}$. 
The induced $S^q$ 
metric ($g_{mn}$) obeys,
\eq{
R_{mn}=\frac{q-1}{L_{S}^2}~\!g_{mn}~, 
}
where $R_{mn}$ is the Ricci tensor of  $S^q$.

Einstein gravity in $(p+q)$ dimensions, minimally coupled to a $q$-form,  
\eq{
S=\int\d^{p+q} x\sqrt{g}\left( R+\frac{1}{2q!}F_{m_1\dots m_q}F^{m_1\dots m_q}\right)
~,}
admits Freund-Rubin solutions \cite{Freund:1980xh} of the form AdS$_p\times M_q$, where 
$M_q$ is a $q$-dimensional Einstein manifold. These solutions obey  \cite{DeWolfe:2001nz},
\eq{\label{a7}
\frac{L_{\text{AdS}}}{L_{\text{int}}}
=\frac{(p-1)}{(q-1)}
~,}
where  $L_{\text{int}}$ is the radius of curvature of $M_q$, 
\eq{
R_{mn}=\frac{q-1}{L_{\text{int}}^2}~\!g_{mn}~, 
}
with $R_{mn}$  the Ricci tensor of  $M_q$. 
In the special case where $M_q$ is the $q$-dimensional sphere, 
$L_{\text{int}}$ reduces to the radius $L_S$.

\section{Spherical harmonics}\label{app:a}

Consider the $q$-dimensional unit sphere $S^q$,
\eq{
S^q=\Big\{
\vec{x}\in\mathbb{R}^{q+1}~|~\sum_{i=1}^{q+1}(x^{i})^2=1
\Big\}
~.}
The basis of spherical harmonics on $S^q$ is inherited from the space of degree-$k$ homogeneous polynomials of $\mathbb{R}^{q+1}$, 
\eq{ p_k(\vec{x})=c_{i_1\dots i_k}
x^{i_1}\dots x^{i_k}~;~~~k\in\mathbb{N}
~,}
where the coefficients are totally symmetric and traceless (for $k\geq2$),
\eq{
c_{i_1\dots i_k}= c_{(i_1\dots i_k)}~;~~~c_{i_1i_2\dots i_k}\delta^{i_1i_2}=0
~.}
Equivalently, these are the polynomials  which are harmonic with respect to 
the scalar Laplacian of $\mathbb{R}^{q+1}$, 
\eq{
\Delta_{\mathbb{R}^{q+1}}p_k(\vec{x})= 
0~.}
Restricting to $S^q$, it follows that they are eigenstates of the scalar Laplacian of $S^q$,\footnote{This can be seen by rewriting the Laplacian of $\mathbb{R}^{q+1}$ in spherical coordinates, 
\eq{
\Delta_{\mathbb{R}^{q+1}}= \frac{\partial^2}{\partial\rho^2}+
\frac{q}{\rho} \frac{\partial}{\partial\rho}+
\frac{1}{\rho^2}\Delta_{S^q}p_k(\vec{x})
~,}
where $\rho$ is the radial coordinate. Taking into account 
that $p_k$ is homogeneous of degree $k$: $\rho\frac{\partial}{\partial\rho} p_k=k p_k$,  and restricting to $\rho=1$,  we obtain \eqref{spheig}.}
\eq{\label{spheig}
\Delta_{S^q}p_k(\vec{x})= 
-k(k+q-1) p_k(\vec{x})
~.}
For a $q$-sphere of radius $L_S$, 
the KK mass-squared is minus the lowest-order non-vanishing Laplacian eigenvalue,
\eq{m_{\text{KK}}^2
=\frac{q}{L_S^2}~.
}
Suppose that orbifolding $S^q$ by a finite subgroup $\Gamma\subset SO(q+1)$ projects out the first $k-1$ 
non-trivial eigenmodes. 
It follows that
\eq{m_{\text{KK}/\Gamma}^2
=\frac{k(k+q-1)}{L_S^2}~,
}
so that from \eqref{1} we obtain the relative order of scale separation,
\eq{\label{a8}
r=\frac{k(k+q-1)}{q}
~.}

\vfill\break

\subsection{Spherical harmonics on $S^2$}\label{app:a1}

We list a basis of the first few harmonic polynomials on $\mathbb{R}^3$. Restricted to the unit sphere $S^2\subset\mathbb{R}^3$ they provide 
a basis of spherical harmonics on $S^2$.
\eq{\spl{\label{a5}
k=0:~&1\\
k=1:~&x,~y,~z\\
k=2:~&xy,~xz,~yz,~x^2-z^2~,y^2-z^2\\
k=3:~&xyz,~x^3-3xy^2,~x^3-3xz^2,~y^3-3yx^2,~y^3-3yz^2,~z^3-3zx^2,~z^3-3zy^2\\
k=4:~&x^3 y - x y^3,
~ y^3 z - y z^3,
~ z^3x-x^3 z
\\
~&x^3 z - 3 x z y^2,
~ y^3 x- 3 x y z^2,
~ z^3 y-3y zx^2,
\\
~&x^4 - 6 x^2 y^2 + y^4,
~y^4 - 6y^2 z^2 + z^4,
~x^4 - 6x^2 z^2 + z^4\\
k=5:~& x^5 - 10 x^3 y^2 + 5 x y^4,
~x^5 - 10 x^3 z^2 + 5 x z^4,
~  y^5 - 10 x^2 y^3+5 x^4 y,
~  y^5 - 10 y^3 z^2 + 5 y z^4,
\\
~&z^5- 10 x^2 z^3  +5 x^4 z,
~ z^5- 10 y^2 z^3 +5 y^4 z ,
~ x^3 y z - x y z^3,
~  x^3 y z - x y^3 z,
\\
~& x^4 y - 6 x^2 y z^2 + y z^4,
~  x y^4 - 6 x y^2 z^2 + x z^4,
~ x^4 z - 6 x^2 y^2 z + y^4 z
\\
k=6:~&
x^6 - 15 x^4 y^2 + 15 x^2 y^4 - y^6,
~x^6 - 15 x^4 z^2 + 15 x^2 z^4 - z^6,
~y^6 - 15 y^4 z^2 + 15 y^2 z^4 - z^6,
\\
~&x^5 z - 10 x^3 y^2 z + 5 x y^4 z,
~ y^5 z- 10 x^2 y^3 z +5 x^4 y z,
~x^5 y - 10 x^3 y z^2 + 5 x y z^4,
\\
~&x y^5 - 10 x y^3 z^2 + 5 x y z^4,
~y z^5 - 10 x^2 y z^3 +5 x^4 y z,
~ x z^5 - 10 x y^2 z^3+5 x y^4 z,
\\
~&3 x^5 y - 10 x^3 y^3 + 3 x y^5, 
~3 x^5 z - 10 x^3 z^3 + 3 x z^5,
~3 y^5 z - 10 y^3 z^3 + 3 y z^5,
\\
~&x^6 - 15 x^4 y^2 + y^6 + 90 x^2 y^2 z^2 - 15 y^4 z^2 - 
 15 x^2 z^4 + z^6
~.}}

\section{Finite subgroups of $SU(2)$}\label{app:su2}

The $SU(2)$ subgroups were classified over a century ago \cite{Klein1888}. They consist of two infinite series and three exceptional cases \cite{slodowy}:

\begin{itemize}

\item The cyclic groups $\mathbb{Z}_{n}$, $n\geq2$, generated by,
\eq{
\left( {\begin{array}{cc}
e^{\frac{2\pi i}{n}} &0\\
  0 & e^{-\frac{2\pi i}{n}} 
   \end{array} } \right)
~,}

\item The binary dihedral groups $\mathbb{D}_{n}$,  obtained by combining $\mathbb{Z}_{n}$ with  the generator $i\sigma^1$.

\item The binary tetrahedral group, obtained  by combining $\mathbb{D}_2$ with the generator, 
\eq{\frac{1}{\sqrt{2}}
\left( \begin{array}{cc}
\varepsilon^7 &\varepsilon^7\\
  \varepsilon^5 & \varepsilon
  \end{array}  \right)
~,}
where $\varepsilon\equiv e^{\frac{\pi i}{4}}$.

\item The binary octahedral group   obtained  by combining the binary tetrahedral 
with, 
\eq{\frac{1}{\sqrt{2}}
\left( \begin{array}{cc}
\varepsilon &0\\
 0 & \varepsilon^7
  \end{array}  \right)
~.}

\item The binary icosahedral group generated by,
\eq{\label{b4}
-\left( {\begin{array}{cc}
\eta^3 &0\\
 0 & \eta^2
  \end{array} } \right)~~\text{and}~~\frac{1}{\eta^2-\eta^3}
\left( {\begin{array}{cc}
\eta+\eta^4 &1\\
  1 & -\eta-\eta^4
  \end{array} } \right)
~,}
where $\eta=e^{\frac{2\i\pi}{5}}$.

\end{itemize}

\section{The icosahedral group}\label{app:ico}

The rotational  (chiral) icosahedral group $\mathcal{I}$ of order 60 
 is isomorphic to ${A}_5$, the alternating group of five elements. 
 $\mathcal{I}$  can be obtained from the binary icosahedral 
subgroup of $SU(2)$, {\it cf.}~\S\ref{app:su2}, via the 2:1 map $SU(2)\rightarrow SO(3)$. 
The preimage in $SU(2)$ of the generator  \eqref{2.2} is the second generator in \eqref{b4}. 
The latter can be put  in the standard $SU(2)$ form,
\eq{ 
\left( {\begin{array}{cc}
a &b\\
  -b^* & a^*
  \end{array} } \right)~;~~~
  a=e^{-\frac{\i}{2}(\alpha+\gamma)}\cos\frac{\beta}{2}~,~b=-e^{-\frac{\i}{2}(\alpha-\gamma)}\sin\frac{\beta}{2}
~,}
with Euler angles: $\sin\frac{\beta}{2}=\tfrac{1}{\sqrt{2}}\sqrt{1+\frac{1}{\sqrt{5}}}$, 
$\cos\frac{\beta}{2}=\tfrac{-1+\sqrt{5}}{\sqrt{10-2\sqrt{5}}}$, $\alpha=0$, $\gamma=\pi$. 
It follows that the image of this element in $SO(3)$ is $R_y(\beta)\cdot R_z(\pi)$, where $\sin\beta=\frac{2}{\sqrt{5}}$, $\cos\beta=-\frac{1}{\sqrt{5}}$. 
%
%
\begin{figure}[tb!]
\begin{center}
\includegraphics[width=0.7\textwidth,trim={0 4cm 0 4cm},clip]{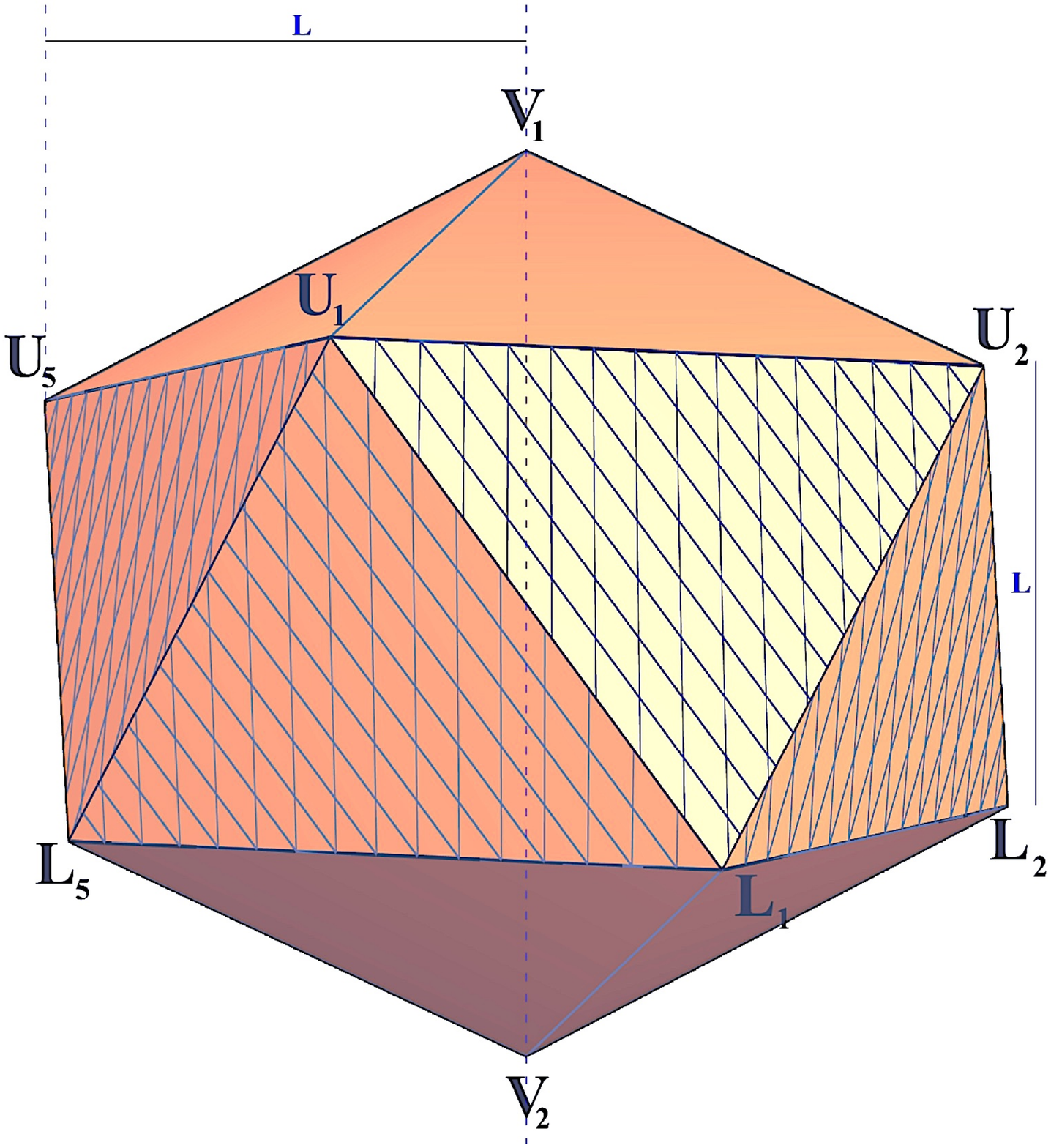}
\caption{The icosahedron. Only eight of the twelve vertices are depicted: the top and bottom vertices ($V_{1,2}$), three of the upper pentagon vertices ($U_{1,2,5}$) and three of the lower pentagon vertices ($L_{1,2,5}$).   
The distance $L$ from each of the vertices of the upper and lower pentagons to the $z$-axis, is also equal to the vertical distance between the upper and lower pentagons. 
}\label{fig:icosahedron}
\end{center}
\end{figure}
 As for any finite group, the
elements of $\mathcal{I}$ can be generated by a set of 
elements satisfying certain relations. This is known as a 
“presentation” of the group. Several different presentations of $\mathcal{I}$ exist with either two or three basis elements \cite{moser}. 
The presentation given in   \cite{shirai}, which  was argued in \cite{Everett:2008et} to be more suitable for flavor model building, uses two elements $S$, $T$ satisfying $S^2=T^5=(T^2ST^3ST^{-1}STST^{-1})^3=\mathbb{I}$. It can be verified that $R_z( \frac{2\pi}{5})$ and the element in \eqref{2.2}  provide an explicit 
representation of $T$, $S$ respectively.

One can also explicitly construct the icosahedron which is invariant under  these generators. In Cartesian coordinates of $\mathbb{R}^3$,  its twelve vertices
are given by: the top and bottom vertices $(0,0,\pm R)$; the upper pentagon vertices $R_z( \frac{2n\pi}{5})\cdot(-L,0,\frac{1}{2}L)$, $n=0,\dots,4$; the lower pentagon vertices  $-R_z( \frac{2(n+1)\pi}{5})\cdot(L,0,\frac{1}{2}L)$. The constants $R=\sin\frac{2\pi}{5}$ and 
$L=\frac{1}{2\sin\frac{\pi}{5}}$, represent respectively the radius of the circumscribing sphere and the distance from each of the vertices of the upper and lower pentagons to the $z$-axis.; the length of the pentagon edges is equal to one. 
It can be seen  that $L$  is also equal to the vertical distance between the upper and lower pentagons. 
Note that the upper and lower pentagons are rotated by $\frac{\pi}{5}$ relatively to each other. Let $V_1$, $V_2$ 
be the top, bottom vertices, $U_1, \dots, U_5$ the vertices of the upper pentagon in the order listed above, and 
$L_1, \dots, L_5$ the vertices of the lower pentagon. To show the invariance of the icosahedron under the group $\mathcal{I}$, it suffices to 
note that the generator $R_z( \frac{2\pi}{5})$ acts on the vertices as the permutation $(V_1,V_2,U_2, \dots, U_5, U_1,L_2, \dots, L_5, L_1)$, 
while the generator \eqref{2.2} acts as the permutation 
$(L_3, U_1,V_2,L_4,U_4,U_3,L_2,L_5,U_5,V_1,U_2,L_1)$.

Let $W$ be any of the 12 vertices   of the icosahedron. 
Thinking of  the group elements of $\mathcal{I}$  as permutations of the 12 vertices,  we shall denote by $g(W)$ 
an element which corresponds to a  permutation of the form $(W,\dots)$. I.e.  $g(W)$ is a rotation which brings the vertex $W$ in the place of the top vertex $V_1$. 
Given 12 elements $g(W)$, $W\in \{V_1,V_2,U_1, \dots, U_5,L_1, \dots, L_5\}$,  the 60 
elements of $\mathcal{I}$  can be obtained from $g(W)$ by successive $\tfrac{2\pi}{5}$ rotations along the vertical axis: $R_z( \frac{2n\pi}{5})\cdot g(W)$, $n=0,\dots,4$. 
Explicitly we take,

\vfill\break

\eq{\spl{\label{explicitrlrmrnts}
g(V_1)&=\mathbb{I}\\
g(V_2)&=RR_zRR_z^{-1}RR_z\\
g(U_1)&=RR_z^{3}\\
g(U_2)&=RR_z^{-1}\\
g(U_3)&=R\\
g(U_4)&=RR_z\\
g(U_5)&=RR_z^2\\
g(L_1)&=RR_zRR_z^{-1}\\
g(L_2)&=RR_zR\\
g(L_3)&=RR_zRR_z\\
g(L_4)&=RR_zRR_zR\\
g(L_5)&=RR_zRR_z^3
~,}}
where $R_z\equiv  R_z( \frac{2\pi}{5})$ and $R$ is the generator of \eqref{2.2}.

\end{appendix}

\bibliography{refs}
\bibliographystyle{unsrt}
\end{document}